# EPITAXIAL GROWTH OF A SILICENE SHEET

Boubekeur Lalmi[a], Hamid Oughaddou[b,c,*], Hanna Enriquez[b,d], Abdelkader Kara[e,c], Sébastien Vizzini[f], Bénidicte Ealet[a], and Bernard Aufray[a,*]

[a]CINaM-CNRS, Campus de Luminy, Case 913, 13288, Marseille, France

[b]CEA, DSM-IRAMIS-SPCSI, Bât. 462, Saclay, 91191, Gif-sur-Yvette, France

[c]Département de Physique, Université de Cergy-Pontoise, 95000, Cergy-Pontoise, France

[d]Département de Physique, Université Paris-Sud, 91405, Orsay Cedex, France

[e]Department of Physics, University of Central Florida, Orlando FL 32816, USA

[f]Aix-Marseille Université, IM2NP, Faculté des Sciences de Saint-Jérôme case 142, 13397 Marseille, France

Abstract

Using atomic resolved scanning tunneling microcppy, we present here the experimental evidence of a silicene sheet (graphene like structure) epitaxially grown on a close-packed silver surface (Ag(111)). This has been achieved via direct condensation of a silicon atomic flux onto the single-crystal substrate in ultra-high vacuum conditions. A highly ordered silicon structure, arranged within a honeycomb lattice is synthesized and presenting two silicon sub-lattices occupying positions at different heights (0.02 nm) indicating possible $sp^2$-$sp^3$ hybridizations.

**Keywords: Silicon, Silicene, Ag, STM, epitaxy, 2D structure**

Silicene presents a honeycomb structure similar to that of graphene, which has demonstrated applications in nanotechnology and is currently the most investigated material in physics and nanoscience[1]. The synthesis of this silicon-based crystalline material opens the way for studies of its physical/chemical properties – some of which have been predicted from theory to be similar to those of graphene – that have potential applications in nanotechnology with the added advantage of being compatible with existing semi-conductor devices. Silicene has recently attracted strong theoretical attention[2-4]. For example, Cahangirov *et al*[4], using density functional theory, have shown that a silicon quasi 2-dimensional structure is stable (without of any imaginary vibrational frequency) only if a low buckling (LB), 0.044 nm, is allowed. They also report that for this LB structure, the Si-Si nearest-neighbour distance is reduced to 0.225 nm from that of the bulk, and that the electronic density of states indicates the system is ambipolar. Their calculated band structures exhibit a crossing at *K* and *K′* points (reflecting the semi-metallic character of silicene) and a linear dependency near the crossing (reflecting a mass-less Dirac fermion character). The same authors find that silicene nano-ribbons (NRs) exhibit electronic and magnetic properties likewise very similar to those of graphene. From an experimental point of view, however, silicene sheets have not been synthesized until now, although single-wall[5] and multi-wall[6] silicon nano-tubes have been produced, and monolayers of silicon have been synthesized by exfoliation[7] (though no 2D character has been detected).

Our group began studying silicon adsorption on silver surfaces several years ago. On Ag(001) we observed super-structures that we eventually came to characterize as deformed hexagonal structures[8]. Subsequently, extensive investigations of Si sub-monolayer growth on Ag(110) revealed self-assembled nanoribbons (NRs) 1.6 nm in width and several hundred nm in length[9-11]. These NRs show strong resistance to oxidation on their edges[12], an intriguing chemical property given the strong propensity of silicon surfaces to oxidize. For these Si

NRs, atomically resolved scanning tunneling microscopy (STM) images revealed a honeycomb structure[11,13] that *ab initio* calculations[14] indicated to be composed of arched silicene NRs. Recently, using angle-resolved photoemission spectroscopy (ARPES), we have inferred that, as with that of graphene, the band dispersion along the direction of these NRs exhibits Dirac cones[15].

The question thus arose as to whether a continuous film of such honeycomb structures – namely silicene – could be achieved.
Very recently we have succeeded in growing a silicene sheet on a Ag(111) surface, in ultra-high vacuum conditions, that we believe is the first evidence of silicene sheet synthesis. The apparatus in which the experiments have been performed is equipped with the standard tools for surface preparation and characterization: an ion gun for surface cleaning, Low Energy Electron Diffraction (LEED) instrument, an Auger Electron Spectrometer (AES) for chemical surface analysis and Scanning Tunneling Microscope (STM) for surface characterization at the atomic scale. The STM experiments were performed at RT with a commercial Omicron STM. The Ag(111) sample was cleaned by several cycles of sputtering (600 eV $Ar^+$ ions, P ~ $10^{-5}$ Torr ) followed by annealing at 400°C until a sharp p(1×1) pattern was obtained. Silicon, evaporated by direct current heating of a piece of Si wafer, was deposited onto the Ag(111) surface held at ~ 250°C. The silver substrate temperature was controlled by a thermocouple located close to the sample.

The experimental conditions for producing this silicon structure are quite stringent since the temperature of the substrate must be kept between 220 and 250°C and the rate of deposition of silicon must be lower than 0.1 monolayer per minute. Finally, extreme care must be taken that *exactly* the amount of silicon required to produce a *single* monolayer is applied. Under these conditions, we produced the silicene sheets in three experiments.

As displayed in Fig. 1a, the atomically resolved STM image shows a highly ordered silicon monolayer, arranged within a honeycomb lattice that covers the entire scanned area of the substrate (several hundred nm$^2$), while Fig. 1b presents an STM image showing that the silicene film covers the surface steps like graphene grown on metals [16,17]. A detailed scrutiny of the STM images revealed that this structure is actually formed by two silicon sub-lattices occupying positions at different heights (Fig. 2a). Analysis of several line-scan profiles – an example of which is shown in Fig. 2b – yields a height difference of 0.02 nm. This small observed corrugation (indicating possible $sp^2$-$sp^3$ hybridizations) is in line with the theoretically predicted one[4].

These line-scan profiles contain several pieces of structural information which can be used to derive the Si-Si nearest-neighbours distance in silicene. A peak-to-peak analysis provides a direct measure of this distance, which is turns out to be between 0.19 and 0.2 nm, while a valley-to-valley analysis indicates a Si-Si distance of about 0.18 to 0.19 nm. That this distance (0.19±0.01 nm) is about 17% shorter than that for Si bulk (0.235 nm) suggests that the Ag substrate may play a catalyst role in the formation and stabilization of the silicene sheet . That the low energy electron diffraction (LEED) of the epitaxially grown film shows a (2√3x2√3)R30° superstructure confirms the existence of a long-range order. In addition the superstructure observed in the LEED pattern is not visible in the STM images probably due to a weak electronic coupling between the silicene sheet and the Ag substrate.

Combining atomic high-resolution STM images recorded on the same sample before and after the silicon deposition without any rotation (shown in Fig. 3a and 3b, respectively) in conformity with the observed (2√3x2√3)R30° LEED pattern, we propose a model of silicene sheet adsorbed on a Ag(111) surface (Fig. 3c). Understanding the growth mechanisms and the stability of the silicene sheet will allow us to assess the viability of silicene as a material with perspectives for potential applications in nanotechnology. Atomistic (*ab initio* and tight-binding) calculations are in progress to unravel the mechanisms by which such a structure is formed. The ARPES measurements are also being programmed in order to reveal the silicene band dispersion and to test the existence of Dirac cones.

**Acknowledgements:**

We thank Lyman Baker for a critical reading of the manuscript. We thank also Guy Le Lay for fruitful discussions. AK thanks the University of Cergy-Pontoise for hospitality and support.

**References:**

[1] A. K. Geim and K. S. Novoselov, Nature Materials **6**, 183 (2007)

[2] S. B. Fagan, R. J. Baierle, R. Mota, Z. J. R. da Silva and A. Fazzio, Phys. Rev. B, **61**, 9994 (2000)

[3] S. Lebègue and O. Eriksson, Phys. Rev. B, **79**, 115409 (2009)

[4] S. Cahangirov, M. Topsakal, E. Aktürk, H. Şahin, and S. Ciraci, Phys. Rev. Lett. **102**, 236804 (2009)

[5] M. De Crescenzi, P. Castrucci, M. Scarselli, M. Diociaiuti, Prajakta S. Chaudhari, C. Balasubramanian, T. M. Bhave, and S. V. Bhoraskar, Appl. Phys. Lett., **86**, 231901 (2005).

[6] S. Yamada and H. Fujiki, Jap. J. Appl. Phys., **45**, L837 (2006).

[7] N. Hideyuki, M. Takuya, H. Masashi, H. Kayo, N. Hiroshi, T. Naoko, N. Takamasa, S.Yoshiki, and N. Hiroshi, Angew. Chem., **118**, 6451 (2006)

[8] C. Léandri, H. Oughaddou, B. Aufray, J M. Gay, G. Le Lay, A. Ranguis and Y. Garreau, Surf. Sci., **601**, 262 (2007)

[9] C. Léandri, G. Le Lay, B. Aufray, C. Girardeaux, J. Avila, M.E. Davila, M.C. Asensio, C. Ottaviani, and A. Cricenti, Surf. Sci., **574**, L9 (2005)

[10] G. Le Lay, B. Aufray, C. Léandri, H. Oughaddou, J.-P. Bibérian, P. De Padova, M.E. Dávila, B. Ealet, A. Kara, Appl. Surf. Sci, **256**, 524 (2009)

[11] A. Kara, C. Léandri, M. E. Dávila, P. De Padova, B. Ealet, H. Oughaddou, B. Aufray and G. Le Lay, J. Supercond. Nov. Magn. **22**, 259 (2009)

[12] P. De Padova, C. Léandri, S.Vizzini, C. Quaresima, P. Perfetti, B. Olivieri, H. Oughaddou, B. Aufray and G. Le Lay, Nano Lett., **8**, 2299 (2008)

[13] B. Aufray, A. Kara, S. Vizzini, H. Oughaddou, C. Léandri, B. Ealet, and G. Le Lay, Appl. Phys. Lett. **96**, 183102 (2010);

[14] A. Kara, S. Vizzini, C. Léandri, B. Ealet, H. Oughaddou, B. Aufray, and G. Le Lay, J. Phys. Condens. Matter., **22**, 045004 (2010)

[15] P. De Padova, C. Quaresima, C. Ottaviani, P. M. Sheverdyaeva, P. Moras, C. Carbone, D. Topwal, B. Olivieri, A. Kara, H. Oughaddou, B. Aufray and G. Le Lay, Appl. Phys. Lett., **96**, 261905 (2010)

[16] E. Loginova, Shu Nie, Konrad Thürmer, Norman C. Bartelt, and Kevin F. McCarty, Phys. Rev. B, **80**, 085430 (2009).

[17] Peter Sutter, Jerzy T. Sadowski, and Eli Sutter, Phys. Rev. B, **80**, 245411 (2009).

**Figure captions:**

**Figure 1:** a) Large-scale, filled-state STM image showing the graphene-like structure of one monolayer of silicon deposited on a close-packed silver surface, Ag(111); b) 3D STM image showing Ag(111) steps covered by the silicene film (11nm x 11nm).

**Figure 2**: a) Filled-state atomically resolved STM image showing honeycomb structure, revealing two sub-lattices; b) Line-profile joining neighboring Si atoms along the direction indicated in Fig. 2a.

**Figure 3:** a) Filled-state atomically resolved STM image of the clean Ag(111) surface. b) Filled-state atomically resolved STM image of the same sample (without any rotation) after deposition of one silicon monolayer. c) Proposed ball model of silicene on Ag(111) derived from both STM images (a) and (b) and from the observed $(2\sqrt{3} \times 2\sqrt{3})R30°$ LEED pattern.

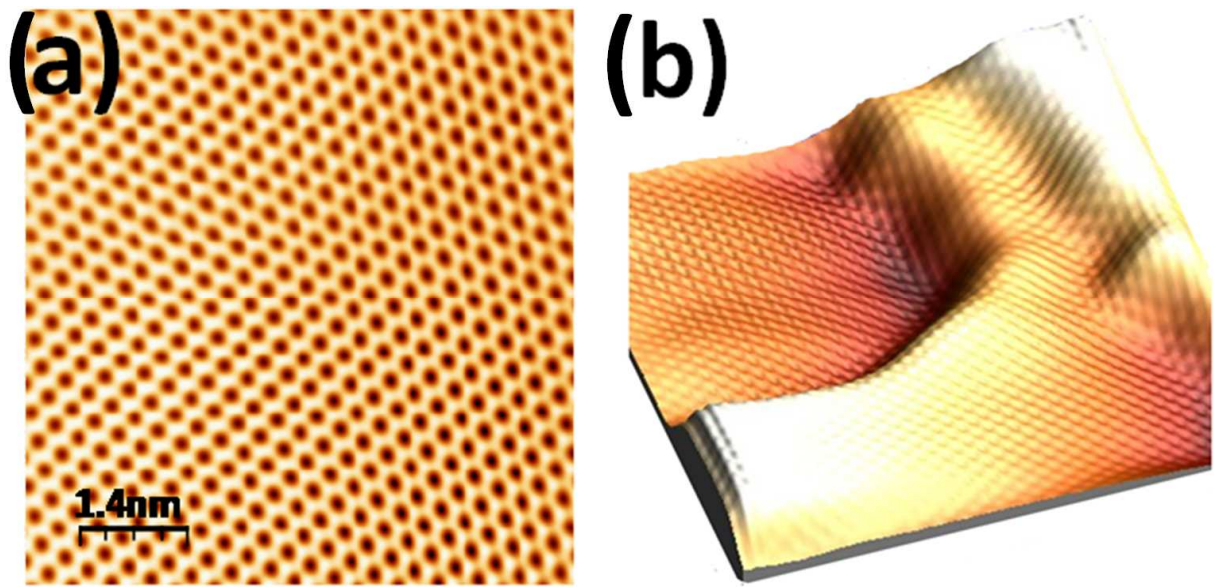

**Figure 1**

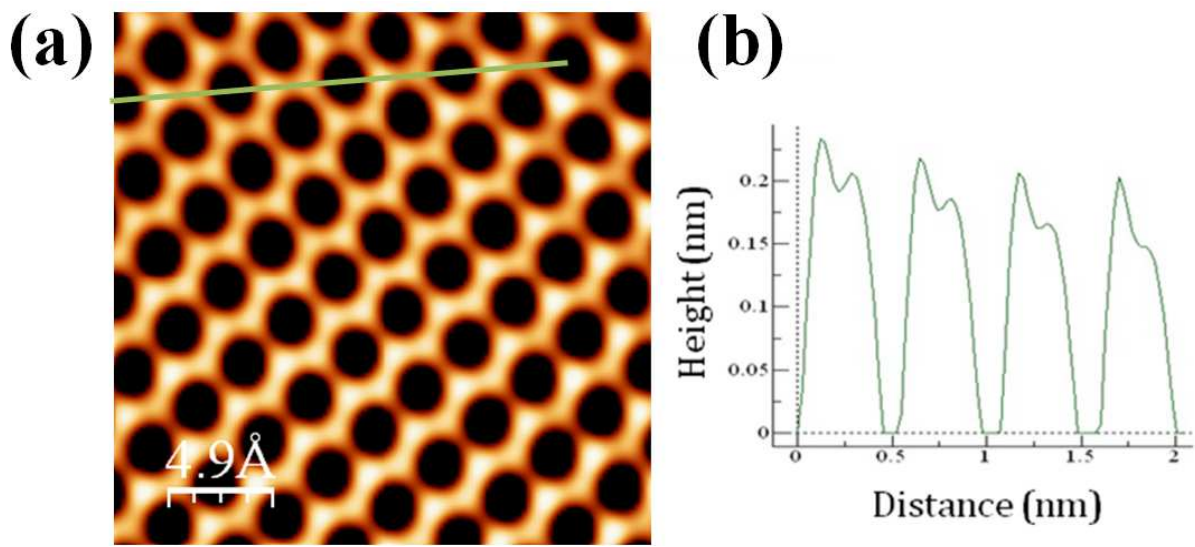

**Figure 2**

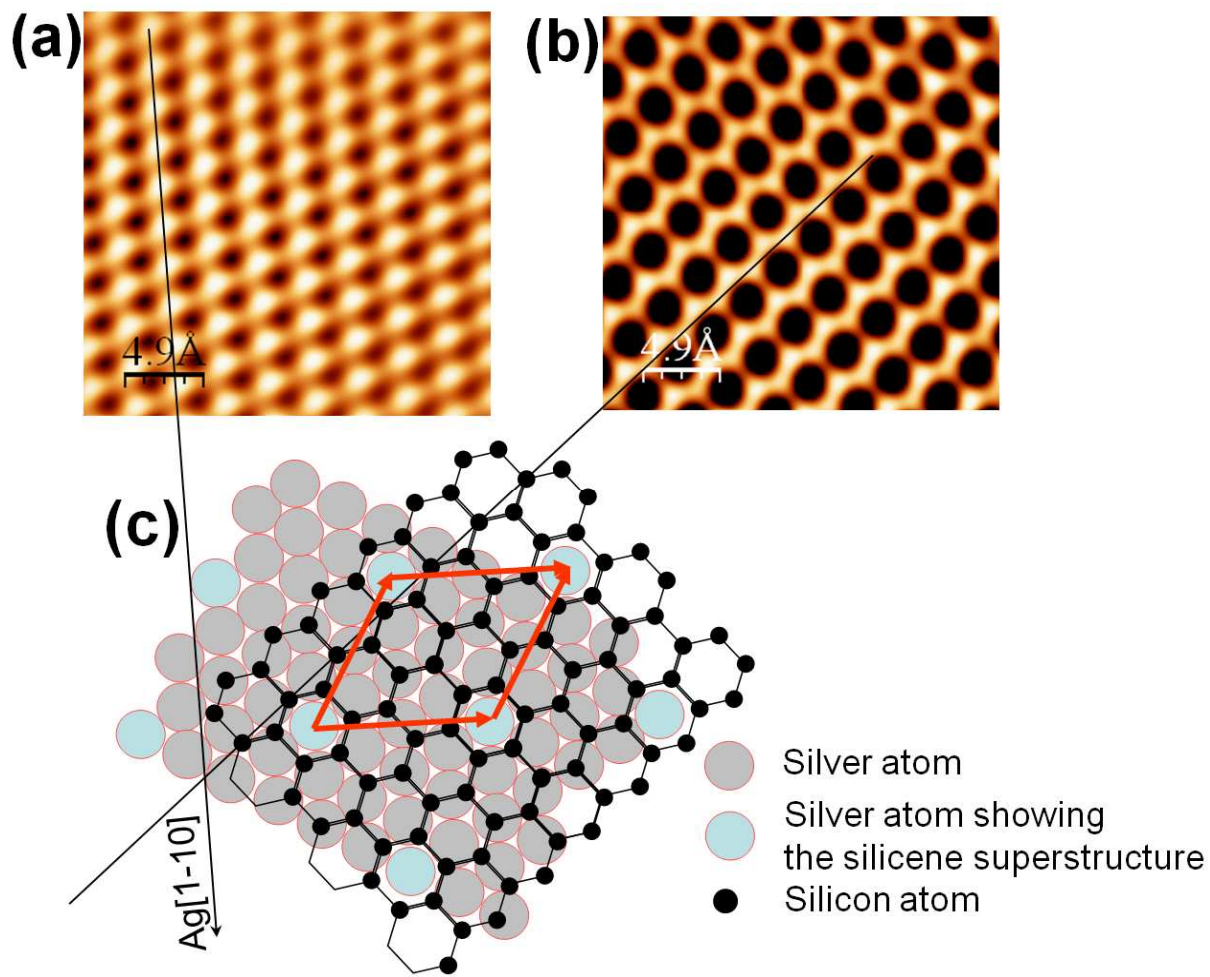

**Figure 3**